\documentclass{scrartcl}

\usepackage{graphics}
\usepackage{multicol}

\begin{document}
\title{CO adsorption on neutral iridium clusters}
%

% \thanks{\emph{Present address:} Insert the address here if needed}%
        % Insert a name or remove this line
%
%\offprints{fielicke@fhi-berlin.mpg.de}

%
% 
%

%end of abstract
%
\author{Christian Kerpal \and Daniel J. Harding  \and 
Gerard Meijer \and Andr\'{e} Fielicke\thanks{e-mail: \texttt{fielicke@fhi-berlin.mpg.de}}}

\maketitle

Fritz-Haber-Institut der Max-Planck-Gesellschaft, Faradayweg 4-6, 14195 Berlin, Germany

\abstract{ The adsorption of carbon monoxide on neutral iridium clusters in the
size range of $n=3$ to 21 atoms is investigated with infrared
multiple photon dissociation spectroscopy. For each cluster size
only a single $\nu$(CO) band is present with frequencies in the
range between 1962\,cm$^{-1}$ ($n=8$) and 1985\,cm$^{-1}$ ($n=18$)
which can be attributed to an atop binding geometry. This behaviour
is compared to the CO binding geometries on clusters of other group
9 and 10 transition metals as well as to that on extended surfaces.
The preference of Ir for atop binding is rationalized by
relativistic effects on the electronic structure of the later 5\textit{d}
metals.} 

\begin{multicols}{2}

\section{Introduction}

 \label{intro} The structure and chemical behaviour of small gas-phase transition
metal (TM)
clusters has been of great interest in the last decade due to their
use as model systems for active sites in heterogeneous catalysis
\cite{Cox1988,SCHNABEL1992,Shi1998,Knickelbein1999}. In particular
the chemistry of carbon monoxide on transition metal clusters has
been intensively studied \cite{Balteanu2003,Socaciu2003,Fielicke2009} because of its relevance in a large number of catalytic
processes. The chemical and physical properties of the clusters
and their complexes can, in principle, be well characterized under
isolated conditions in the gas phase. Thus they offer good benchmark
systems for quantum chemical calculations, which become particularly
challenging for the late 5\textit{d} transition metals. The ''CO on Pt(111)
puzzle'' is a well-known example for the difficulties in correctly
predicting CO adsorption sites by density functional theory (DFT),
see below \cite{Feibelman2001,Hu2007}. The availability of
structural information for cluster complexes is therefore vital for testing and
improving theoretical methods.

% , as well as information about the influence of
% relativistic effects on the chemical properties of the cluster
% \textbf{Mentioning of relativistic effects does not fit here.}
% \cite{Fielicke2006,Gruene2008a,Lyon2009}.

% \textbf{There is a big jump to the next sentence. Why is vibrational
% spectroscopy helpful for this? give refernce also to older IR-REMPI
% work, see Knickelbein. Your refs are only for work with FELIX. refer
% to Chem. Phys. Lett. 357, 195 (2002). the following senctence is too
% general}

% Recently the technique of  (IR-MPD) has been successfully applied
% for this purpose via measuring the vibrational spectra of
% \emph{naked and/or decorated} \textbf{would you understand this??}
% clusters in several studies
% \cite{Fielicke2004,Swart2007,Gruene2008,Fielicke2009}.

Carbon monoxide is a ligand particularly suited for such
investigations. The character of the M--CO bond is
commonly described within the Blyholder model \cite{BLYHOLDER1964} of $\sigma$-donation and
$\pi$-backdonation. The strength of the internal C--O bond and
consequently the C--O stretching frequency is very sensitive to the degree of donation/backdonation and hence to the
binding geometry and the electronic structure of the cluster.
Infrared multiple photon dissociation (IR-MPD) spectroscopy allows
the measurement of cluster size and composition specific IR-spectra
of CO-complexes of transition metal clusters in the gas phase to
determine, e.g., the CO binding geometries. Thereby it has been found
that, in general, at low coverage CO binds to clusters of the 3\textit{d}
transition metals only in atop-configuration while for 4\textit{d} and 5\textit{d}
transition metals bridging and face-capping CO ligands can also be
present \cite{Fielicke2009}. For instance, in a comparative study of the group 10 transition
metals (Ni, Pd, Pt) it was found that Ni clusters bind CO only in
atop positions while Pd clusters show a great variety of binding
sites \cite{Gruene2008a}. Pt clusters again show only atop binding,
which was rationalised by the increasing role relativistic effects
play in determining the electronic structure for heavier elements,
leading to a direct influence of these effects on the chemical
properties. This direct influence has been observed previously for
Pt(100) surfaces by Pacchioni \textit{et al.} \cite{Pacchioni1997}
using density functional theory calculations. Their results show a
drastic increase in the adsorption energies for both atop and bridge
bound CO when relativistic effects were taken into account. However,
since the effect was stronger for the atop bound CO the energy
ordering of the binding geometries changed, with atop binding
now being favoured over bridge binding. A number of other DFT
studies of the CO adsorption on Pt surfaces using Local Density
Approximation (LDA) and General Gradient Approximation (GGA) failed
to predict the correct absorption site on the Pt(111) surface,
namely atop, in contrast to the calculated hollow site preference (see \cite{Feibelman2001} and references therein).
It has been shown that a better treatment of both relativistic effects and of the electronic structure, especially the exchange correlation, are important and can help to solve the puzzle \cite{Hu2007,Orita2004,Doll2004,Stroppa2008}. However, even with current DFT methods such calculations remain a challenging subject with the need for further benchmark data, preferably on elements which are strongly influenced by relativistic effects.

%Orita \textit{et al.} got the right site-prefernence with an all electron scalar relativistic calculation, while Doll compared the results of PW91 and B3LYP both with a
%scalar-relativistic pseudopotential. He found that the B3LYP hybrid functinal predicted the correct binding site, while PW91 failed. Stroppa \textit{et al.} predicted the correct binding sites for Cu and Rh but failed for Pt using the hybrid functinals HSE03 and PBE0 (both with a scalar-relativistic pseudopotential). However, a comparison between PBE and PBE0 showed a destabilization of the hollow site using the hybrid functional. So even with up to date DFT methods the calculations remain a challenging subject with the need for further benchmark data preferably on systems which are strongly influenced by reltivistic effects.

One of these elements is iridium, as it is adjacent to platinum in the 5\textit{d} metals. Here we
present our recent findings on the binding geometry of CO
adsorbed on neutral Ir clusters in the size rage of 3 to 21 atoms,
thus complementing our previous studies on the group 9 transition
metals Co and Rh \cite{Fielicke2006}.

% As discussed below the group 9 metals show a very similar behaviour
% compared to the group 10 metals: The clusters of the 3d metal Co
% bind CO only in atop position, Rh clusters have a variety of binding
% sites. Ir clusters, again like Pt clusters, bind CO only in atop
% position. For a full review of CO binding on charged and neutral
% transition metal clusters see Fielicke \textit{et al.}
% \cite{Fielicke2009}.

\section{Experimental Techniques}
\label{sec:1} IR-MPD spectra are obtained by irradiating a molecular beam containing Ir$_{n}$CO complexes
with intense IR radiation and measuring the changes induced in the
mass distribution. At a vibrational resonance, a cluster complex can absorb (multiple) IR photons and,
if the absorbed energy is sufficient, dissociation can be induced. IR
spectra are constructed by analyzing the intensity changes as a
function of the IR wavelength.

All experiments reported here are performed at the Free Electron
Laser for Infrared eXperiments (FELIX) facility \cite{OEPTS1995} in
the Netherlands, and the details of the experiments have been
reported elsewhere \cite{Fielicke2003,Lyon2009}.
Briefly described, iridium clusters are generated by laser ablation of a rotating Ir rod. A continuous
stream of He serves as carrier gas, while a mixture of 1 \% CO in He
is introduced 60 mm downstream the ablation in a flow
reactor \cite{Fielicke2000}. The level of CO is
adjusted such, that the clusters adsorb at most a single CO
molecule. The cluster beam expands into vacuum where it is
irradiated by the counterpropagating FELIX beam. The charged
particles are deflected from the molecular beam while the neutral
clusters continue to propagate until they are ionized by 7.9 eV
photons from an F$_2$ excimer laser in the acceleration region of a
reflectron time-of-flight mass spectrometer. The molecular beam
experiment runs at 10\,Hz with a repetition rate of FELIX of 5\,Hz
so that mass spectra with and without FELIX radiation are recorded
alternately, accounting for changes in cluster intensity. The IR
region covered is 1580\,cm$^{-1}$ to 2040\,cm$^{-1}$ corresponding
to the typical frequencies of the C--O stretches for the different
binding geometries.

\section{Results and Discussion}
\label{sec:2} 
\begin{figure}
\resizebox{1.00\columnwidth}{!}{%
  \includegraphics{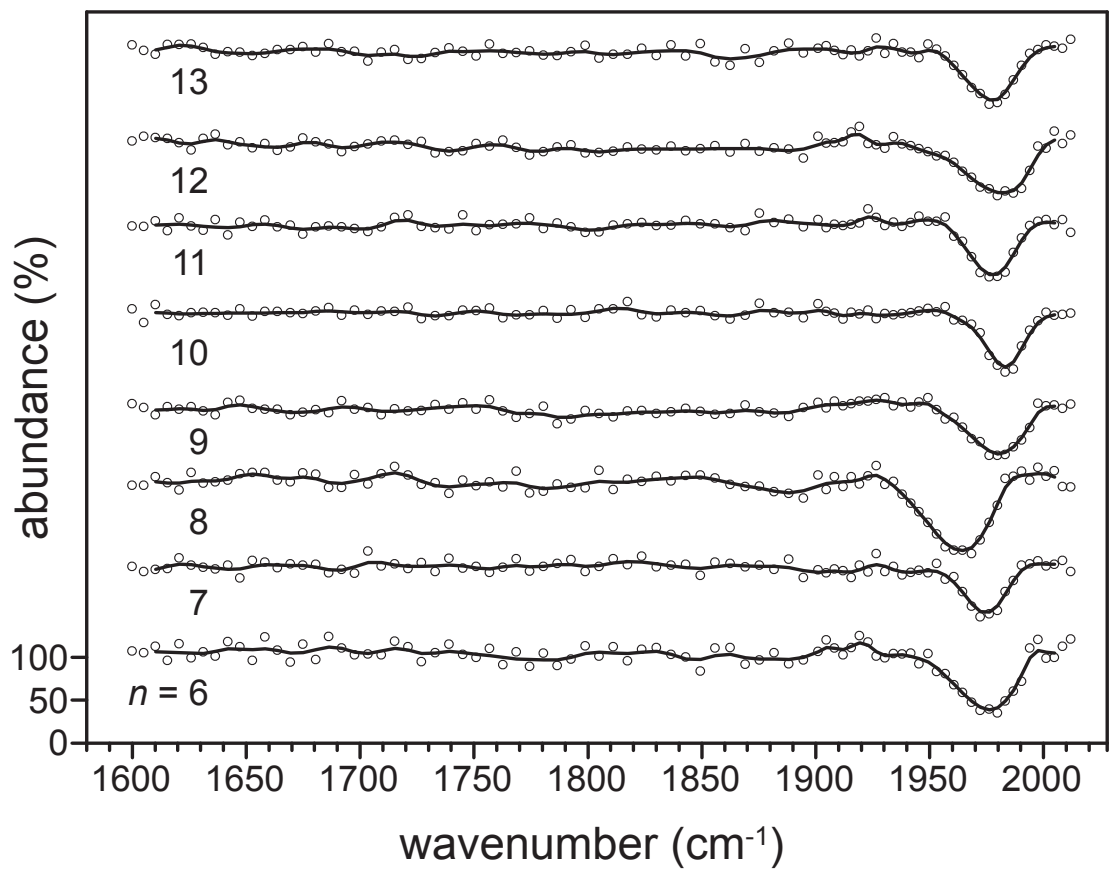}
}
%\vspace{5cm}       % Give the correct figure height in cm
\caption{IR-MPD spectra of CO adsorbed on neutral Ir clusters of
different size \textit{n}. The dots are the actual data points,
the line is a binomially weighted five point average.}
\label{fig:1}       % Give a unique label
\end{figure}

\begin{figure}
\resizebox{1.00\columnwidth}{!}{%
  \includegraphics{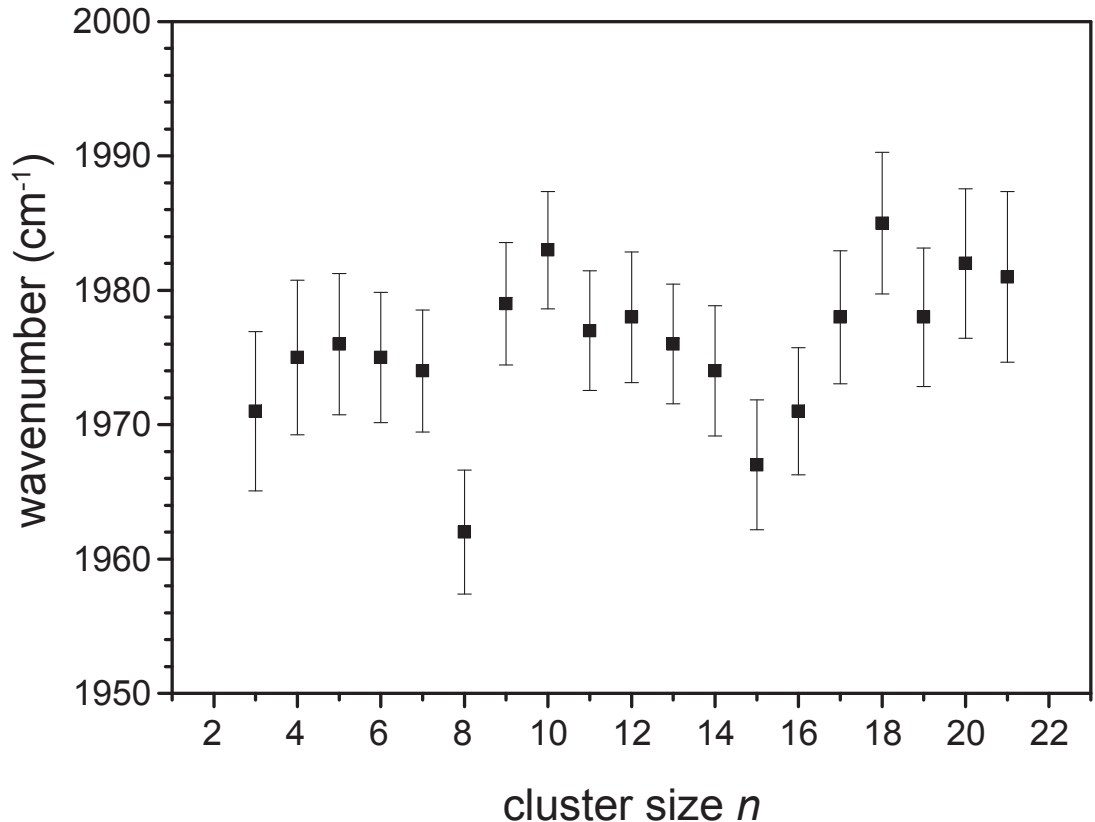}
}
%\vspace{5cm}       % Give the correct figure height in cm
\caption{Peak position of the $\nu$(CO) band for different cluster sizes \textit{n}. The error stems from two parts: an uncertainty in the wavelength calibration of about 0.2 $\%$ that equally applies to all values (thus not changing relative peak positions) and the standard deviation of the least square fit to a Gaussian line shape function used to determine the peak positions ($\leq 2.5$\,cm$^{-1}$).} 
\label{fig:2}       % Give a unique label
\end{figure}

The depletion spectra of the Ir$_n$CO complexes for $n=6$ to 13 are shown in
figure \ref{fig:1}. Similar spectra have been measured for the range from $n=3$ to 21 and the dependence of the frequencies on
the cluster size is shown in figure \ref{fig:2}. The peak positions of the
$\nu$(CO) bands are determined by a least-squares fit to a Gaussian
line shape function. In the  1580--2040\,cm$^{-1}$ range only a
single band is present, which can be attributed to the internal $\nu$(CO) stretch of an atop, $\mu^1$,
bound CO molecule as discussed below. The band position varies smoothly with cluster size, with the exception of $n=8$. The reason for the shift in the frequency of more than 10\,cm$^{-1}$ compared to the values for $n=7$ and $n=9$ is unclear, but may be related to a change in geometry, as binding to lower coordinated metal atoms is expected to lead to a red-shift of the $\nu$(CO) stretching frequency (\textit{vide infra}).

\end{multicols}

\begin{multicols}{2}

The assignment of the $\nu$(CO) bands to an atop binding geometry
is made by comparison with previous measurements of CO molecules
bound to TM surfaces \cite{MARTIN1993,Lauterbach1996} and clusters
\cite{Fielicke2006,Lyon2009}. For the neutral Ir clusters being
discussed here, the frequencies of $\nu$(CO) are at
minimum 1962\,cm$^{-1}$ and at maximum 1985\,cm$^{-1}$. This
corresponds to a red-shift of about 40-60\,cm$^{-1}$ compared to the
values reported for atop bound CO on Ir(100) and Ir(111) surfaces
for low coverage, 2026\,cm$^{-1}$ and 2030\,cm$^{-1}$, respectively
\cite{MARTIN1993,Lauterbach1996}. Such a shift is consistent with
results for other transition metal clusters
\cite{Gruene2008a,Fielicke2006,Lyon2009}. For example, the red-shift
for atop bound CO on Pt clusters compared to different Pt surfaces
is about 50-80\,cm$^{-1}$ \cite{Gruene2008a}, representing the
general finding of a red-shift of the CO stretching frequencies for
all neutral (and anionic) transition metal clusters compared to the
extended surface values. The stronger C--O bond activation on the
cluster can be explained by the lower coordination of the metal atoms, similar to the effect that has been shown both experimentally \cite{Klunker1996} and theoretically \cite{BRANDT1992,Curulla1999} for CO adsorption on different Pt surfaces, where lower coordination numbers for the Pt atom binding the CO molecule lead to larger red shifts of the C--O stretching frequency. A possible explanation is related to the $\pi$-backdonation occurring from the interaction of the filled metal \textit{d}-orbitals with the CO
$\pi^*$-orbital. This effect is expected to be stronger for the lower coordinated clusters atoms. However, Curulla \textit{et al.} calculate a similar contribution for both the $\sigma$-donation and
$\pi$-backdonation regardless of the coordination number, suggesting a different reason, e.g. differences in substrate polarization or differences in Pauli repulsion \cite{Curulla1999}.  

CO binds to extended Ir surfaces only in atop positions, even at higher coverages \cite{Lauterbach1996,Landolt,Titmuss2002}. However, Gajdo\u{s} \textit{et al.} calculated frequencies for higher coordinated CO binding sites to be in a range from \textit{ca.} 1730 to 1825\,cm$^{-1}$ \cite{Gajdos2004}, while also calculating $\nu$(CO) to be 2041\,cm$^{-1}$ for the atop binding geometry. There is a difference of between 240 and 150\,cm$^{-1}$ for the $\nu$(CO) stretching frequency for these higher coordinated sites, compared to our values. In contrast, the difference for the atop binding geometry is only about 65\,cm$^{-1}$, supporting our assignment, even without taking into account the red-shift of the  $\nu$(CO) bands for TM clusters compared to extended surfaces.  

Higher coordinated CO ligands are present, however, in the saturated cluster carbonyl Ir$_{6}$(CO)$_{16}$ that exists in two isomers, a black one with four $\mu$$^{2}$--CO ligands and a red one with four $\mu$$^{3}$--CO ligands. The $\nu$(CO) frequencies for their vibrations are \textit{ca.} 1840 and 1760\,cm$^{-1}$, respectively, significantly lower than our observations \cite{GARLASCHELLI1984}. 

Zhou \textit{et al.} measured CO stretching frequencies for IrCO in neon matrices \cite{Zhou1999}. Their value of 2024.5\,cm$^{-1}$ is about 50\,cm$^{-1}$ higher in energy compared to our mean value of 1976\,cm$^{-1}$. The same trend is true when comparing their results for CoCO and RhCO with our earlier measurements on Co$_{n}$CO and Rh$_{n}$CO. The CO stretching frequencies in the neon matrices for Co and Rh are about 44 and 62\,cm$^{-1}$ higher in energy, respectively \cite{Fielicke2006,Zhou1999}. It is possible that these differences can be attributed to matrix effects since even the choice of the matrix element can change the frequency significantly (e.g. for RhCO from 2022.5\,cm$^{-1}$ in neon to 2007.6\,cm$^{-1}$ in argon).

%More details on the binding mechanism can be found in References
%\cite{Fielicke2009,Fielicke2006,Pacchioni1997,Gajdos2004}.
The assignment to atop bound CO is also consistent with DFT
calculations (UB3LYP) on Ir$_{13}$ clusters with CO adsorbed by
Okumura \textit{et al.} \cite{Okumura2006}. They include scalar
relativistic effects to calculate the interaction of Ir$_{13}$ with
CO and other small ligands, assuming a cuboctahedral geometry for
the cluster. As they calculate a C--O stretching frequency of 2190\,cm$^{-1}$ for gas phase C--O, while the real value is 2143\,cm$^{-1}$, their results need to be scaled for a comparison. This leads to a vibrational frequency of 1978\,cm$^{-1}$ for atop bound CO (unscaled 2021\,cm$^{-1}$) which is in good agreement with our experimental value of
1976\,cm$^{-1}$.

\end{multicols}

\begin{multicols}{2}

As mentioned above, to our knowledge, CO is bound to Ir surfaces only in atop positions \cite{Lauterbach1996,Landolt,Titmuss2002}. In contrast to
the case of Pt, previous DFT calculations have consistently predicted
the experimentally observed binding site (see \cite{German2008} and references therein). However, there exists no direct comparison of calculations with and
without inclusion of relativistic effects as is the case for Pt.
As the group 9 metal clusters show a very similar behaviour compared
to the group 10 metal clusters, namely atop binding for the 3\textit{d}
metals, various binding sites for 4\textit{d} metals and again atop binding
for the 5\textit{d} metals, a common reason for the change in CO adsorption
geometry seems plausible. That is, that relativistic effects lead to
a contraction and decrease in energy for the \textit{s}- and
\textit{p}-orbitals while \textit{d}- and \textit{f}-orbitals
increase in energy due to a radial expansion, with the effect being
biggest for the 5\textit{d} metals, moderate for the 4\textit{d} metals and almost
negligible for the 3\textit{d} metals. This results in an ordering of the
M--CO bond length for the group 10 metals that is Ni \textless \,Pt
\textless \,Pd whereas without relativistic influences the Pt--CO
bond length would be the largest. As this decrease in bond length is
large for atop binding but almost nonexistent for bridge binding, the
former is strongly stabilized, leading to an unusual preference for
this binding site for the late 5\textit{d} metals \cite{Pacchioni1997}. It is
interesting to compare this to the case of tungsten and rhenium. Both are 5\textit{d}
metals but, in contrast to Pt and Ir, their clusters bind CO in both atop positions and higher coordination sites \cite{Lyon2009}. This behaviour was
rationalized by a smaller relative stabilization of the atop site
compared to the late 5\textit{d} TMs due to the spatially more extended
5\textit{d} orbitals.

\section{Conclusion}
\label{summ}

The adsorption of CO on neutral Ir clusters in the size range of 3 to
21 atoms has been investigated with IR-MPD spectroscopy.
%between
%1580\,cm$^{-1}$ to 2040\,cm$^{-1}$ corresponding to the possible
%vibrational frequencies of the CO stretches for the different
%binding geometries.
The vibrational spectra reveal that the only CO-binding geometry
present is atop binding with $\nu$(CO) frequencies between 1962\,cm$^{-1}$ and 1985\,cm$^{-1}$.
%The assignment of that geometry was
%done by comparison to the CO binding properties on various
%transition metal clusters and surfaces.
These results complement earlier studies on the other group 9 and 10
TM clusters showing a similar behaviour for both groups with respect
to the observed CO-binding geometries. The 3\textit{d} metals bind CO only in
atop position, the 4\textit{d} metals have various (size dependent) binding
sites and the 5\textit{d} metals again show only atop binding. The behaviour
of the 5\textit{d} metals can be explained by the increased influence of
relativistic effects on their electronic structure.
%As a consequence the M-CO atop bond is considerably shortened and thus strongly
%stabilized.
Interestingly, the earlier 5\textit{d} TMs tungsten and rhenium do not show the preference
of platinum and iridium because of their spatially more extended 5\textit{d}
orbitals and the consequently smaller stabilisation of the atop
binding.
With our measurements we provide further benchmark data on a system that is strongly affected by relativistic effects, thus aiding in testing and developing appropriate theoretical models for the challenging calculations on TM clusters.

%\begin{acknowledgment}
\thanks{
We gratefully acknowledge the support of the Stichting voor
Fundamenteel Onderzoek der Materie (FOM) in providing beam time on
FELIX. We thank the FELIX staff for their skillful assistance, in
particular Dr. A.F.G. van der Meer and Dr. B. Redlich. This work is
supported by the Deutsche Forschungsgemeinschaft through research
grant AF 893/3-1 and the Cluster of Excellence UNICAT hosted by the
Technical University Berlin. D.J.H. acknowledges support from the
Alexander-von-Humboldt-Stiftung.}

%\end{acknowledgment}

\
%\begin{figure*}
%\vspace*{5cm}       % Give the correct figure height in cm
%\end{figure*}
%
% BibTeX users please use
\end{multicols}

\bibliographystyle{epj}

\end{document}